\documentclass[twocolumn,showpacs,preprintnumbers,amsmath,amssymb]{revtex4}

\usepackage{graphicx} 
\usepackage{dcolumn}  
\usepackage{bm}       

\begin{document}

\title{Interaction-Induced Localization of an Impurity 
in a Trapped Bose Condensate}

\author{Ryan M. Kalas}

\author{D. Blume}

\affiliation{%
Department of Physics and Astronomy, Washington State University, Pullman, 
Washington 99164-2814, USA}


\date{December 1, 2005}

\begin{abstract}
We study the ground state properties of a trapped Bose condensate with a neutral 
impurity.  By varying the strength of the attractive atom-impurity interactions 
the degree of localization of the impurity at the trap center can be controlled.
As the impurity becomes more strongly localized
the peak condensate density, which can be monitored experimentally, grows markedly.
For strong enough attraction, the impurity can make the condensate unstable 
by strongly deforming the atom density in the neighborhood of the impurity.
This ``collapse'' can possibly be investigated in bosenova-type experiments.

\end{abstract}

\maketitle


\section{Introduction}
The study of impurities immersed in liquids and 
solids has a long history. In 1933, Landau
predicted, using quantum mechanical arguments, that 
the localization of electron impurities
in a crystal
could be used to probe the activation energy of solids~\cite{land33}.
Electron impurities have also played a key role
in the study of liquids, in particular liquid $^4$He~\cite{eHereview}.
More recently, the study of doped mesoscopic helium clusters
has attracted much attention~\cite{toen01,call01}. 
Some atom impurities reside on the cluster
surface while others migrate to the center of the helium cluster.
Spectroscopic measurements of molecules located at the
center of the cluster have, e.g., shown unambiguously 
that $^4$He clusters with about 60 atoms are superfluid~\cite{greb98}.

Recently, the study of impurities immersed in a gaseous, coherent atom
background has become possible~\cite{chik00,ciam02}. 
Theoretical studies on, e.g., ion impurities in a condensate have been
initiated~\cite{cote02,mass05}, raising questions about the
appropriate treatment of systems with long-range
interactions 
(unlike short-ranged atom-atom potentials, which behave as $1/r^6$
for large interparticle distances, atom-ion potentials fall off as $1/r^4$).
Here, we consider a neutral impurity in an inhomogeneous Bose 
gas, assuming contact atom-impurity interactions~\cite{cucc}.
Treatments for more complicated
atom-impurity interactions exist~\cite{capu00,nave99,chin00}; the results may,
however, be model-dependent.
Our self-consistent mean-field
treatment provides a first step towards a systematic understanding of
impurities in a Bose condensate.  We also discuss a simple 
variational treatment which reproduces the key features
of the self-consistent results.  We point towards possible experimental
signatures of our predictions, which will be aided by the possibility of tuning
the atom-atom and atom-impurity interactions in the vicinity of a Feshbach resonance
by application of an external magnetic field~\cite{inou98,corn00}.
This tunability is unique to gaseous condensate-impurity systems; it does not,
for example, exist in helium where the interaction strength is set by nature.

\begin{figure}
\includegraphics[scale=0.53]{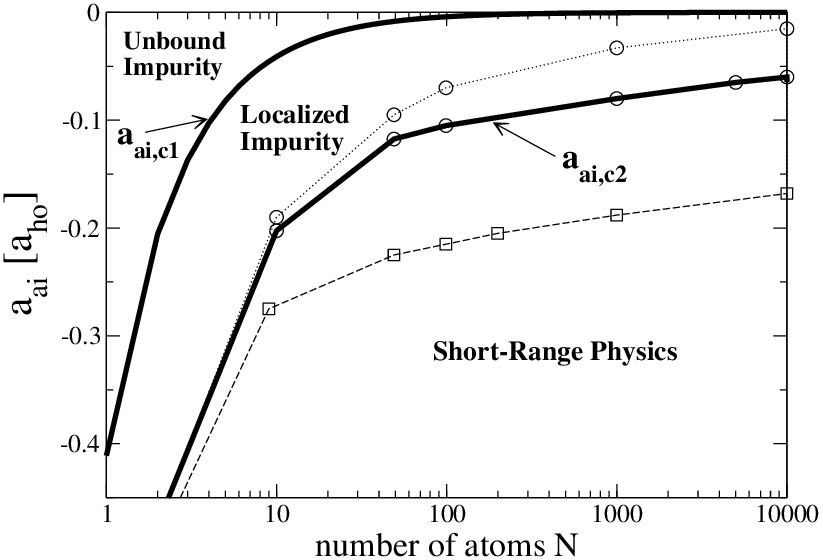}
\caption{\label{phase} 
Phase diagram for trapped Bose gas with a single impurity,
which feels no confining potential,
as a function of the number of atoms $N$ and the atom-impurity
scattering length $a_{ai}$ for equal atom and impurity mass, i.e.,
$m_i=m_a$.  The phase diagram 
contains three regions: in region (A) the impurity
is unbound; in region (B) the impurity is localized
(the localization is ``weak'' for comparitively small $|a_{ai}|$
and ``strong'' for comparatively large $|a_{ai}|$, see Sec.~\ref{meanfield} 
for details); and in region (C) short-ranged
physics becomes relevant.  Regions (A) and (B) are separated by a critical 
value $a_{ai,c1}$ 
(upper bold solid line), which is approximately independent of $a_{aa}$.
Regions (B) and (C) are separated by a critical value $a_{ai,c2}$, 
which is shown for $a_{aa}=0.005a_{ho}$ (lower bold
solid line), $a_{aa}=0$ (dotted line), and $a_{aa}=0.05a_{ho}$ (dashed line).}
\end{figure}

We consider a weakly-interacting Bose condensate in a harmonic trap, 
doped with a single impurity.
For now, we assume that the impurity feels no external trapping
potential; later, we discuss how the presence of an impurity
trapping potential modifies the results.
Figure 1 shows the equilibrium ``phase diagram''~\cite{footnote_phase} 
determined within mean-field theory as a function of the
number of atoms $N$ and the atom-impurity scattering length $a_{ai}$.
The phase diagram separates into three distinct regions:
(A) For $a_{ai}>a_{ai,c1}$, the impurity is unbound and can move away
from the trapped atom cloud.
(B) For $a_{ai,c1}>a_{ai}>a_{ai,c2}$, the impurity is 
localized, i.e., bound to the atom cloud~\cite{footnote_chem}.
(C) For $a_{ai}<a_{ai,c2}$, short-range physics, which cannot
be described within mean-field theory, becomes relevant.
Regions (A) and (B) are separated by a $N$-dependent
critical value $a_{ai,c1}$ (upper solid bold line in Fig.~1), which is
approximately independent of 
the atom-atom scattering length $a_{aa}$.
Since the impurity feels no trapping potential, interaction-induced localization of the impurity 
occurs only if $a_{ai}$ is more attractive 
than $a_{ai,c1}$.
Regions (B) and (C) are separated by a $N$-dependent
critical value $a_{ai,c2}$, which also depends on the atom-atom
scattering length $a_{aa}$. 
The lower bold solid line in Fig.~1 shows
$a_{ai,c2}$ for $a_{aa}=0.005 a_{ho}$,  
the dotted line that for $a_{aa}=0$, 
and the dashed line that for $a_{aa}=0.05a_{ho}$.  
If $a_{ai}$ is more negative than $a_{ai,c2}$, 
the attractive atom-impurity interactions can ``collapse'' the condensate,
pulling atoms into a short-ranged state about the impurity.

The next section outlines the self-consistent mean-field treatment
used to calculate the phase diagram shown in Fig.~1.
Section~\ref{variational} 
develops a simple variational framework, which reproduces
the key features of the full self-consistent mean-field
treatment.
Finally, Sec.~\ref{conclusion}
discusses possible experimental realizations of the systems under
study and concludes.

\section{Self-consistent mean-field treatment}
\label{meanfield}

We describe $N$ atoms 
\begin{figure}
\includegraphics[scale=0.5]{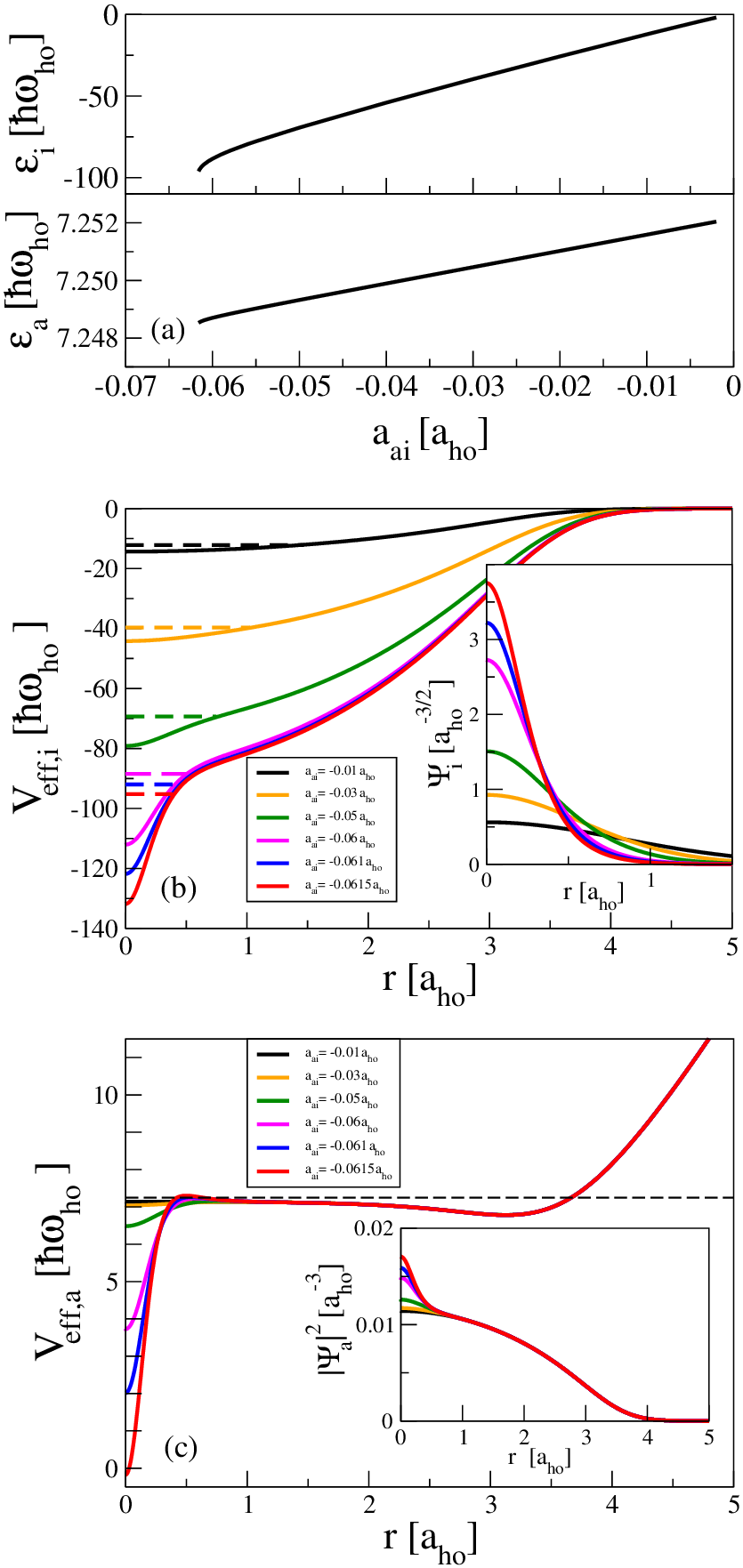}
\caption{\label{Attr_fig} 
(color online)
Self-consistent mean-field results obtained for attractive 
atom-impurity interactions, $N=10^4$, $a_{aa}=0.005a_{ho}$, and $m_i=m_a$.
Panel (a) shows the chemical potentials $\epsilon_i$ and $\epsilon_a$ 
as a function of $a_{ai}$.
Solid lines in panels~(b) and (c) show the effective potentials
$V_{eff,i}(r)$ and $V_{eff,a}(r)$, respectively, for a few selected 
atom-impurity scattering lengths (see legend).
Dashed lines show the corresponding chemical potentials
[note that the change of $\epsilon_a$ is not visible on the scale
chosen in panel~(c)].
The insets of panels (b) and (c) show the corresponding impurity wave function
$\psi_i(r)$ and atom density $|\psi_a(r)|^2$, respectively.
The critical values $a_{ai,c1}$ and $a_{ai,c2}$ of this system 
are $\approx -4 \times 10^{-5} a_{ho}$ and $-0.062 a_{ho}$, respectively.  }
\end{figure}
of mass $m_a$ in the presence of a harmonic trapping potential
with angular frequency $\omega_{ho}$ and a single 
impurity of mass $m_i$, which feels
no external potential, within mean-field theory.  Assuming that the atom-atom and atom-impurity interactions
can be described by contact potentials, the many-body Hamiltonian reads 
\begin{eqnarray}
\label{Hmanybody}
H &=&\sum_{j=1}^{N} \left[-\frac{\hbar^2}{2m_a}\nabla_j^2+\frac{1}{2}m_a\omega_{ho}^2 \vec{x}_j^2 \right] 
-\frac{\hbar^2}{2m_i}\nabla_i^2 \nonumber\\
& &+ U_{aa}\sum_{j<k}^{N} \delta(\vec{x}_j-\vec{x}_k) + U_{ai}\sum_{j=1}^{N} \delta(\vec{x}_j-\vec{x}_i), 
\end{eqnarray}
where $U_{qp}=2\pi\hbar^2 a_{qp}/m_{qp}$, $m_{qp}=m_q m_p / (m_q + m_p)$,
and $(q,p)=(a,a)$ or $(a,i)$.
In Eq.~(\ref{Hmanybody}), $\vec{x}_j$ and $\vec{x}_i$ denote
the position vectors
of the $j$th atom and the impurity, respectively.
We approximate the 
ground state wave function $\Phi$ as 
a product of single-particle wavefunctions,
\begin{equation}
\label{ansatz}
\Phi(\vec{x}_1,\vec{x}_2,\cdots,\vec{x}_N;\vec{x}_i) = 
\left[\prod_{j=1}^N \psi_a(\vec{x}_j)\right]
 \psi_i(\vec{x}_i),
\end{equation}
and derive a set of coupled Hartree-Fock equations,
\begin{eqnarray}
\label{HFa}
 \Large[-\frac{1}{2}\frac{\partial^2}{\partial r^2}+ \frac{1}{2}r^2 + (N-1)\frac{a_{aa}}{a_{ho}}
 \frac{|\chi_a(r)|^2}{r^2}+ \nonumber \\
\frac{a_{ai}}{a_{ho}}\frac{m_a}{2m_{ai}}
\frac{|\chi_i(r)|^2}{r^2} \Large]\chi_a(r) 
 =
\epsilon_a \chi_a(r)    
\end{eqnarray}
and
\begin{eqnarray}
\label{HFi}
 \left[-\frac{1}{2}\frac{m_a}{m_i}\frac{\partial^2}{\partial r^2} + N\frac{a_{ai}}{a_{ho}}
 \frac{m_a}{2m_{ai}} \frac{|\chi_a(r)|^2}{r^2} \right]\chi_i(r)=
\epsilon_i \chi_i(r).    
\end{eqnarray}
Here,
$a_{ho}$ denotes the oscillator length ($a_{ho}=\sqrt{\hbar/m_a \omega_{ho}}$
and $\vec{r}=a_{ho}\vec{x}$), 
and $\epsilon_a$ and $\epsilon_i$ 
the chemical potentials (or ``orbital energies'') 
of the atoms and the impurity.
The coupled mean-field equations are equivalent to 
those for a two-component condensate~\cite{TBEC} if one replaces
one of the two components
by a single impurity.
In writing Eqs.~(\ref{HFa}) and (\ref{HFi}), we
have implied spherical symmetry, 
$\psi_{a,i}(\vec{r})=\psi_{a,i}(r)=\frac{\chi_{a,i}(r)}{\sqrt{4\pi}r}$
with $\int_{0}^{\infty} |\chi_{a,i}(r)|^2dr =1$.  
For repulsive atom-impurity interactions, not considered here,
symmetry-breaking states can exist~\cite{SSB}.

The impurity feels an effective potential $V_{eff,i}$   
[defined as the second term in square brackets on the left hand side (LHS) 
of Eq.~(\ref{HFi})], which is created by 
the atom density $|\psi_a|^2$.  
The impurity density $|\psi_i|^2$ enters Eq.~(\ref{HFa}) and creates, 
together with the trapping potential and the atom density itself, an
effective
potential $V_{eff,a}$ [defined as the last three terms in square 
brackets on the LHS of Eq.~(\ref{HFa})].
For weak atom-impurity interactions, the condensate atoms act
to a good approximation as a static
background with which the impurity interacts.  
However, as the strength of the atom-impurity 
interactions increases the full coupled nature of Eqs.~(\ref{HFa})
and (\ref{HFi}) becomes important.


We discuss the self-consistent solutions to Eqs.~(\ref{HFa})
and (\ref{HFi}), obtained by numerical means, for a specific set of parameters.
The behavior is qualititatively similar for other parameters.
Figure~\ref{Attr_fig}(a) shows the chemical potentials 
$\epsilon_i$ and $\epsilon_a$ as a function of $a_{ai}$ 
for $N=10^4$ and $a_{aa}=0.005a_{ho}$, and $m_i=m_a$. 
Equal atom and impurity masses can be realized experimentally by,
e.g., promoting a single condensate atom to a different hyperfine state~\cite{chik00}.
Figure~\ref{Attr_fig}(a) shows that the chemical potentials
$\epsilon_i$ and $\epsilon_a$ change 
approximately linearly with the atom-impurity scattering length for 
$-0.005\gtrsim a_{ai}/a_{ho} \gtrsim -0.05$.  
This linear behavior is what one would expect from a perturbative treatment.  
To visualize the system's behavior, 
Figs.~\ref{Attr_fig}(b) and (c) show the effective potentials
$V_{eff,i}(r)$ and $V_{eff,a}(r)$ for a few selected atom-impurity scattering lengths.
Figure~2(b) shows that $V_{eff,i}$ becomes deeper as $a_{ai}/a_{ho}$
goes from $-0.01$ to $-0.03$ to $-0.05$.  Accordingly,
the impurity wave functions $\psi_i$, shown
in the inset of Fig.~2(b), become more localized as $|a_{ai}|$
increases.  Although $|\psi_i|^2$ changes significantly as $a_{ai}/a_{ho}$ goes from 
$-0.01$ to $-0.05$, $V_{eff,a}$ 
and $|\psi_a|^2$ change only slightly [see Fig.~\ref{Attr_fig}(c)].
 
For $a_{ai}/a_{ho}\lesssim -0.05$,
the impurity chemical potential $\epsilon_i$ (and, to a lesser degree,
the atom chemical potential $\epsilon_a$) changes in a non-linear,
i.e., non-perturbative, fashion.  For the parameters at play here,
this defines the regime of strong atom-impurity coupling.  
To highlight the dramatic changes of the system in this strongly-coupled regime,
Figs.~2(b) and (c) show self-consistent effective potentials 
for three nearly identical atom-impurity scattering lengths,
i.e., $a_{ai}/a_{ho}=-0.06,-0.061,-0.0615$. 
The peak impurity density grows with increasing $|a_{ai}|$ 
and creates a ``hole'' at the center of $V_{eff,a}$, which in turn causes
the atom density to grow a ``bump'' at the trap center
[see inset of Fig.~2(c)] 
with a length scale of roughly the condensate healing length $\xi$.
The healing length is given by the competition 
between the kinetic energy and the condensate's mean-field energy,
$\xi=1/\sqrt{8\pi\rho_a a_{aa}}$ \cite{dalf99}, 
where $\rho_a$ denotes the peak density of the atoms,
$\rho_a=N|\psi_a(r=0)|^2$.  For the parameters of Fig.~\ref{Attr_fig},
$\xi\approx0.26a_{ho}$.  Since the healing length $\xi$ 
is the scale over which the condensate ``reacts'' 
to spatial perturbations, it is natural that the 
atom density develops a variation near the trap center of size $\xi$.  

\begin{figure}
\hspace{-0.3in}
\includegraphics[scale=0.25]{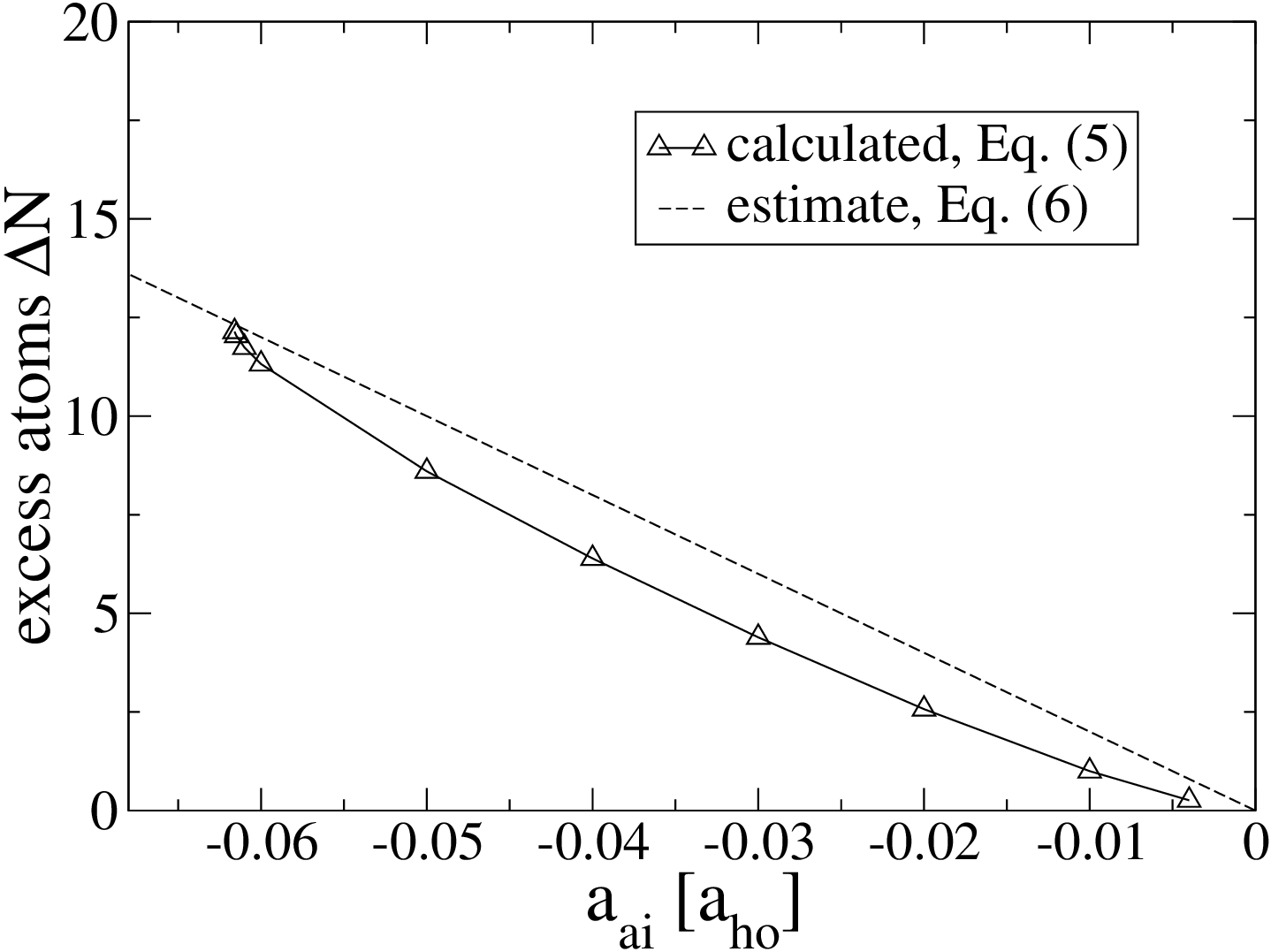}
\caption{\label{fig_excess} 
Triangles show the number of excess atoms $\Delta N$ calculated from
Eq.~(\protect\ref{eq_excdef}) using the self-consistent 
mean-field solutions for
$N=10^4$, $a_{aa}=0.005a_{ho}$ and $m_a=m_i$ (the same parameters as 
in Fig.~2),
as a function of the atom-impurity scattering length $a_{ai}$.
For comparison, a dashed line shows the analytical
estimate, Eq.~(\protect\ref{eq_excess}).}
\end{figure}

The inset of Fig.~\ref{Attr_fig}(c) illustrates the peak atom density
growth with increasing $|a_{ai}|$. To quantify this growth, 
we calculate the excess number of atoms $\Delta N$ associated with the
bump of the atom density.
In analogy to a homogeneous system~\cite{mass05}, 
we define $\Delta N$ as 
\begin{eqnarray}
\label{eq_excdef}
\Delta N = 4 \pi N \int_{0}^{r_c} 
[ |\psi_a^{a_{ai} \ne 0}(r )|^2- |\psi_a^{a_{ai}=0}(r)|^2] r^2 dr,
\end{eqnarray}
where the atom wave function $\psi_a^{a_{ai} \ne 0}(r)$ is
calculated self-consistently for a Bose gas with finite atom-impurity 
scattering length and 
$\psi_a^{a_{ai}=0}(r)$ for a Bose gas with vanishing atom-impurity 
scattering length (for the same $N$ and $m_a$).
When evaluating Eq.~(\ref{eq_excdef}) for a specific system,
we choose the cutoff radius $r_c$ to roughly coincide with the $r$-value
at which the bump of the atom density starts growing. 
Triangles in Fig.~\ref{fig_excess} 
show the resulting number of excess atoms $\Delta N$ for 
$N=10^4$, $m_i=m_a$, $a_{aa}=0.005a_{ho}$ (the same parameters
as in Fig.~\ref{Attr_fig}) and $r_c=1 a_{ho}$ for different
values of the atom-impurity scattering length $a_{ai}$.
The number of excess atoms increases roughly
linearly with increasing $|a_{ai}|$.
Just before the onset of instability at $a_{ai,c2}\approx -0.062a_{ho}$, 
the number of excess atoms $\Delta N$ reaches 12, 
which corresponds to 0.12~\% of the total number of atoms.
For comparison, a dashed line in Fig.~\ref{fig_excess} shows 
an estimate for the number of excess atoms 
derived for a weakly-interacting impurity-doped 
homogeneous Bose gas~\cite{mass05},
\begin{eqnarray}
\label{eq_excess}
\Delta N = - \frac{m_{aa}}{m_{ai}} \frac{a_{ai}}{a_{aa}}.
\end{eqnarray}
Figure~\ref{fig_excess} shows good agreement between 
the number of excess atoms $\Delta N$ calculated for
the inhomogeneous impurity-doped condensate (triangles) and the 
analytical expression, Eq.~(\ref{eq_excess}).
This suggests that Eq.~(\ref{eq_excess}) describes 
the number of excess atoms for large enough, weakly-interacting
inhomogeneous condensates quite accurately.

Finally, if $a_{ai}$ becomes more negative than a critical value of
$a_{ai,c2} \approx -0.062 a_{ho}$,
we no longer find a self-consistent solution to 
Eqs.~(\ref{HFa}) and (\ref{HFi}).
This implies that the condensate collapses,
i.e., atoms are drawn into a short-ranged state about the impurity.
It appears likely that this collapse involves only a fraction
of the condensate atoms, but a definite answer lies
beyond the scope of the present work.
Just as in the case of pure atomic condensates with 
negative atom-atom scattering length,
mean-field theory predicts the onset of collapse for our coupled
equations but cannot describe 
the system's behavior in the regime dominated by 
short-range physics.

We now estimate the critical value $a_{ai,c1}$, which
separates the unbound impurity phase from the localized impurity
phase for $m_i=m_a$ (see upper bold solid line in Fig.~1).
The impurity feels a strictly short-ranged potential, i.e., $V_{eff,i}(r)$
falls off faster than a power-law of $r$.  
Since the impurity equation, Eq.~(\ref{HFi}), is linear,
we can compare the volume-integrated strength of $V_{eff,i}$
with the corresponding critical value, 
$-\frac{\pi^3}{6}\frac{\hbar^2 b}{m_i}$~\cite{Landau},
for forming a bound state in three dimensions in a potential of 
range $b$.  If we identify the range $b$ with $a_{ho}$, 
we find that the critical value of $a_{ai}$
scales as $1/N$, i.e., $a_{ai,c1}(N)/a_{ho}=
-\frac{\pi^2}{24 N}\approx -\frac{0.411}{N}$.  
For positive $a_{aa}$, the
atom cloud is somewhat larger, and our estimate will be off
by a numerical factor of order $1$.
The upper bold solid line in Fig.~\ref{phase} shows our analytical estimate.
The results of our numerical calculations are consistent with this analytical estimate.
The critical value $a_{ai,c1}$ might be difficult to observe 
experimentally since the transition from region (A) to region (B) 
involves a diverging length scale.
Furthermore, it might be difficult to experimentally realize a 
trapping setup with tunable atom-impurity scattering length
$a_{ai}$ for which the impurity feels no confining potential 
(see also Sec.~\ref{conclusion}).

The behavior of the impurity-doped condensate was illustrated in Fig.~2
for $N=10^4$ atoms.  We find similar qualitative features for a smaller
number of atoms, including the
disappearance of the mean-field solutions.  The critical values
$a_{ai,c1}$ and $a_{ai,c2}$ vary with $N$ as shown in Fig.~1.  We note
that the notion of a condensate healing length, and thus
the discussion of the disappearance of the mean-field solution at the point when
the impurity becomes more tightly localized than this scale,
becomes less meaningful for small enough number of atoms.  
It will be interesting to further investigate the properties of an impurity immersed
in a small condensate since such systems can be realized experimentally
with the aid of optical lattices.

\section{Variational treatment}
\label{variational}
To further illustrate how
the neutral impurity
alters the atom cloud,
we minimize the total energy variationally
for the Hamiltonian given in
Eq.~(\ref{Hmanybody}) and the wave function
given in Eq.~(\ref{ansatz}) with
\begin{eqnarray}
\label{psivar}
\psi_i \propto  e^{-p_i r^2}, \hspace{0.18in}\text{and}\hspace{0.18in}     
\psi_a \propto  (e^{-p_a r^2}+ c\hspace{0.03in}e^{-p_b r^2}).
\end{eqnarray}
The variational parameters $p_i$ and $p_a$ determine respectively the width
of the impurity and of the atom wave function.
To be able to describe the growth of the atom peak density 
in the strongly-interacting regime
and
the collapse of the condensate, 
$\psi_a$ contains an additional gaussian with two more variational parameters,
the relative amplitude $c$ and parameter $p_b$.  We restrict $p_b$ to be
greater than $p_a$ to separate the background condensate cloud from
the more localized condensate bump.

\begin{figure}
\hspace{-0.3in}
\includegraphics[scale=0.5]{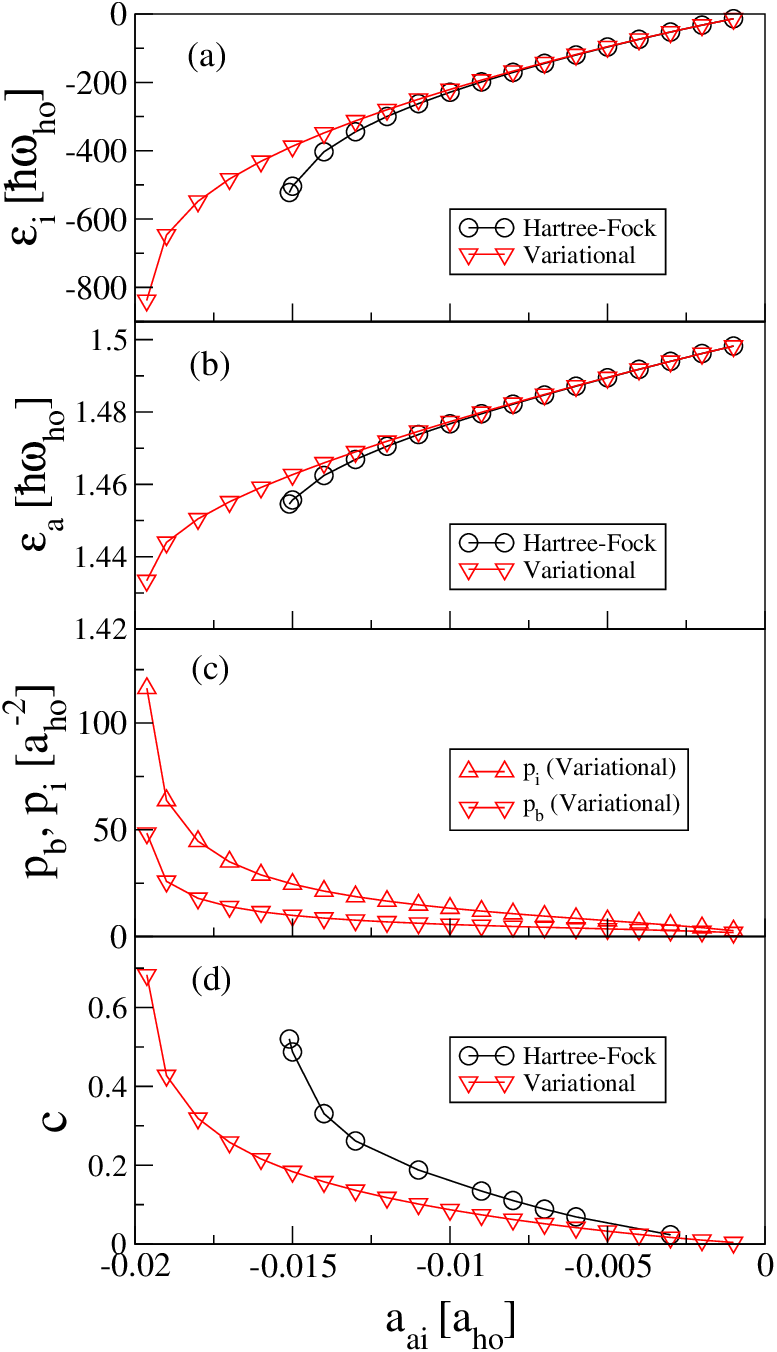}
\caption{\label{fig4} 
(color online)
Triangles show the chemical potentials $\epsilon_i$
and $\epsilon_a$ [panels (a) and (b)],
and the parameters $p_b$, $p_i$ and $c$ [panels (c) and (d)] 
for $N=10^4$, $a_{aa}=0$ and $m_i=m_a$ as a function of
$a_{ai}$, obtained from the variational treatment.
For comparison, circles in panels (a) and (b) show $\epsilon_i$ and $\epsilon_a$,
and those in panel~(d) the parameter $c$ obtained by fitting
the self-consistent solutions of Eqs.~(\protect\ref{HFa}) and (\protect\ref{HFi}) (see text).}
\end{figure}

Figure~\ref{fig4} shows the results of our variational calculations 
for $N=10^4$, $a_{aa}=0$, and $m_i=m_a$ as a function of $a_{ai}$.
To reduce the parameter space we set $p_a=0.5a_{ho}^{-2}$; 
we checked that allowing $p_a$ to vary changes its value only little.
The optimal values of the remaining three variational
parameters $p_i$, $p_b$ and $c$
are shown in panels (c) and (d) of Fig.~4 by triangles,
and the corresponding chemical potentials of the impurity
and atom, respectively, are shown in panels (a) and (b) by triangles. 
The variational analysis 
predicts a critical value of $a_{ai,c2} \approx-0.020 a_{ho}$; at this
critical value $a_{ai,c2}$, the local minimum in the variational energy
disappears as the variational energy becomes unbounded from below.
For comparison, our self-consistent solutions 
to Eqs.~(\ref{HFa}) and (\ref{HFi}),
which are shown in panels (a) and (b) by circles for comparison,
predict a somewhat less attractive critical value, i.e.,
$a_{ai,c2}\approx -0.016a_{ho}$.  
This is to be expected since the self-consistent total energy (not shown)
provides a better lower bound than the variational energy.
The variational parameters $p_i$, $p_b$ and $c$
shown in panels (c) and (d) of Fig.~4 nicely illustrate
the degree of impurity localization. For small
$|a_{ai}|$, the amplitude $c$ is negligible, 
indicating that the presence of the impurity barely affects the condensate.
As $|a_{ai}|$ increases, 
the impurity becomes more tightly localized, i.e., $p_i$ increases
(note that the width of the impurity density scales as $1/\sqrt{p_i}$), which
in turn drives the growth and localization
of the condensate bump, i.e., $c$ and $p_b$ also increase.  
Since $p_i$ drives the increase of $p_b$, 
$p_b$ necessarily increases slower than 
$p_i$ with increasing $|a_{ai}|$.

To connect our results for the variational parameter $c$ with the
full self-consistent solutions, we fit our solutions to Eqs.~(\ref{HFa}) and (\ref{HFi})
to the wavefunctions of Eq.~(\ref{psivar}) with the proper normalization, treating
$p_i$, $p_b$, and $c$ as fitting parameters.  The circles plotted in Fig.~4(d) show the 
resulting values of $c$ extracted from the self-consistent solution. 
To a very good approximation, $c^2$ describes the percentage change in the 
peak condensate density for the system with non-vanishing $a_{ai}$ as compared 
to the system with vanishing
$a_{ai}$ (assuming we keep $a_{aa}$ and the number of atoms $N$ fixed).
Figure~4(d) shows that, just before collapse at $a_{ai,c2}$,
$c \approx 0.53$ for the full self-consistent solution to Eqs.~(\ref{HFa}) and (\ref{HFi}) 
and $c \approx 0.68$ for the variational solution.
These values of $c$ correspond to changes in the peak condensate density, 
as compared to the condensate without impurity, of greater than 25\%.
We note that a similar growth of the peak condensate density is seen in the inset of Fig.~2(c) for the 
same number of atoms, i.e., $N=10^4$, but non-vanishing
atom-atom interactions, i.e., $a_{aa}=0.05a_{ho}$.

The variational wavefunction given in Eq.~(\ref{psivar})
is best-suited to describe the case of $a_{aa}=0$.  For ${a_{aa}\neq}0$ the atom cloud
deviates from a gaussian, and for strong enough interactions a gaussian form
for the atom cloud is a poor approximation.  
Consequently, 
as the atom-atom interactions increase, we find that the simplistic variational wave function 
given in Eq.~(\ref{psivar}) cannot describe the tightly-localized 
impurity at the trap center prior to the onset of collapse.  
Nonetheless, for the case $a_{aa}=0$ discussed above, the variational treatment 
reproduces the key features of the full self-consistent solution and provides 
us with further insights.  In particular, the form of the variational wavefunction, 
Eq.~(\ref{psivar}), is useful in visualizing how the condensate develops features characterized
by a length scale much smaller than the oscillator length.
Furthermore, the disappearance of the local minimum as the variational energy becomes
unbounded from below is another indication, along with the disappearance
of the self-consistent solutions, of the collapse of the condensate.
 
\section{Discussion and conclusion}
\label{conclusion}
Sections~\ref{meanfield} and \ref{variational}
discuss the behaviors of a single neutral impurity, which feels no
external confining potential, immersed in a trapped condensate.
If the impurity feels an external trapping potential with angular frequency $\omega_i$, 
which might be the case in an experiment (see below), 
region (A) in Fig.~1 is absent, i.e., the impurity is always 
localized due to the presence of the external potential.
Assuming that the impurity trapping potential has a characteristic length
which is larger than roughly the condensate healing length (for large 
enough number of atoms), the comparatively strong impurity localization prior 
to collapse and the crossover from region (B) to region (C) in Fig.~1 are, however, 
nearly unaltered.  For example, for $N=10^4$, $a_{aa}=0.005a_{ho}$, $m_i=m_a$ 
and $\omega_a=\omega_i$, the onset of collapse occurs at the 
same critical value of $a_{ai,c2}\approx-0.062a_{ho}$ that we find without impurity trap.  

Impurity-doped condensates can be realized experimentally 
with present-day technology~\cite{chik00}.
If one considers a magnetically trapped condensate, 
an impurity can, e.g., be created by promoting one of the condensate atoms
to a different hyperfine state. The promoted atom may or may not feel
the magnetic confinement. Alternatively, one could
implant a different atom, magnetic or non-magnetic, into the cloud.
Such systems have the disadvantage that the atom-impurity
interactions cannot be tuned via a magnetic Feshbach resonance.
To take advantage of the tunability of interspecies scattering 
lengths~\cite{jin},
one can consider
an optical potential red-detuned with respect to the atoms and the impurity.
In such an experimental realization both the atom and the impurity
would feel trapping potentials.

As the atom-impurity interactions are tuned closer
to $a_{ai,c2}$, the growth of the peak atom density at the center
of the trap can potentially be monitored experimentally in
expansion experiments. Since the condensate bump at the trap center
involves only a few atoms, direct detection of the changes in the peak density 
may, however, be non-trivial.  We suggest that the impurity-doped condensate 
could alternatively be probed in a bosenova-type experiment which applies a 
sequence of time-dependent magnetic field ramps~\cite{bosenova}.
By tuning the atom-impurity scattering length to a large negative
value, one could experimentally induce collapse and consequently
density oscillations, which might involve a significant fraction 
of the condensate atoms.

A key result of our study is that the degree of localization 
of the impurity at the trap center in region (B) of the phase diagram
(see Figs.~\ref{phase} and \ref{Attr_fig})
can be controlled by varying the atom-impurity scattering length, i.e.,
the width of the impurity wave function for a system with
$m_a=m_i$ can be varied from a size much greater than the oscillator length $a_{ho}$ 
to a size significantly smaller than $a_{ho}$.
In addition to changing the atom-impurity scattering length,
one can consider unequal atom and impurity masses, e.g., a Cs
atom immersed in a Na condensate.
Not surprisingly, as the impurity mass increases, 
the degree of impurity localization also increases.
The localized impurity itself may present the possibility 
for forming interesting single-atom devices, perhaps using the impurity's 
spin degrees of freedom.  The favorable coherence properties of Bose condensates 
may make the localized impurity states viable for quantum computing schemes.  
Furthermore, extensions to two or more impurities will
allow one to consider the role of condensate-mediated interactions between impurities.

Finally, we return to our finding 
that a single neutral impurity can deform the condensate sufficiently
to induce a collapse which may only involve a fraction of the condensate atoms.
The resulting collapsed state may be related 
to the mesoscopic droplets that have been predicted
to form about an ion immersed in a condensate~\cite{cote02}.  
More work is needed to fully understand these ion states, the collapsed states
predicted in the present work, and possible connections between
the two. 

We gratefully acknowledge fruitful discussions with 
A. Bulgac, M. Davis, P. Engels, B. Esry and, E. Timmermans, 
hospitality of the Institute for Nuclear Theory
at the University of Washington,
and support by the NSF, grant PHY-0331529.


\begin{thebibliography}{10}
\bibitem{footnote_phase}
The transition from unbound to localized impurity
is associated with a diverging length scale, i.e., it
is a localization transition.  The transition from
the localized impurity to short-range physics is to
be taken figuratively; the critical line
$a_{ai,c2}(N)$ indicates a change of behavior of
the system whose description goes beyond mean-field
theory, as discussed further in the text.
\bibitem{footnote_chem}
Throughout this paper, a localized impurity is characterized by a 
negative chemical potential $\epsilon_i$ of the impurity.
An unbound impurity has a positive chemical potential.
\end{thebibliography}
\end{document}